\documentstyle{article}

\def\abstract#1{\vskip 7mm
        \begin{center}{}\par \smallskip
                \begin{minipage}[c]{12cm}
                        \small #1
                \end{minipage}
        \end{center}
}
\def\title#1{\begin{flushleft}{\Large\bf #1}\end{flushleft}}
\def\author#1{\vskip 5mm \begin{flushleft}{#1}\end{flushleft}}
\def\address#1{\begin{flushleft}{\it #1}\end{flushleft}}

\begin{document}

\setlength\arraycolsep{2pt}

\title{The Kretschmann scalar for 5D Vacua}

\author{Kayll Lake\footnote{E-mail: \tt{lake@astro.queensu.ca}}}

\address{Department of Physics, Queen's University\\
Kingston, Ontario, K7L 3N6, Canada}

\abstract{We present a streamlined calculation of the Kretschmann scalar for 5D Vacua considered recently by Fukui \textit{et al }.
}

\section*{{\large I.~INTRODUCTION}}
Recently, Fukui \textit{et al }$^1$ examined the 5D metric
\begin{equation}
ds^2=\sigma
\{dt^2-a^2[\frac{dr^2}{(1-kr^2)}+r^2d\Omega^2]\}+\epsilon b^2dl^2, \label{metric}
\end{equation}
where $k=\pm1,0$, $d\Omega^2\equiv d\theta^2+\sin^2\theta \, d\phi^2$, $\epsilon=\pm1$, $\sigma=\sigma(l)$, $a=a(t,l)$ and $b=b(t,l)$, with regard to the 5D vacuum condition $R_{AB}=0$. The authors comment that the associated 5D Kretschmann curvature invariant $K\equiv R^{ABCD}R_{ABCD}$ is difficult to calculate even by computer. The purpose of this note is to show how to trivialize this calculation.

\section*{{\large II.~NOTATION AND KRETSCHMANN DECOMPOSITION }}
Write $R_{ABCD}$ for the Riemann tenor, $R_{AB}$ for the Ricci tensor, $R$ for the Ricci scalar ($\equiv R_{A}^{A}$) and $C_{ABCD}$ for the Weyl tensor where the dimension of the space is $n$. A straightforward calculation gives the identity
\begin{equation}
R_{ABCD}R^{ABCD} = \frac{8}{(n-2)^2}R_{AB}R^{AB}+\frac{4(n^2-8n+13)}{(n-1)^2(n-2)^2}R^2+C_{ABCD}C^{ABCD} \label{identity}
\end{equation}
so that for vacua we need consider only the Weyl invariant $C_{ABCD}C^{ABCD} $ which, for the metric (\ref{metric}), is given by

\begin{equation}
\frac{(2\,a a_{ll}  b ^{3}\epsilon-2\, a_{l} ^{2} b ^{3}\epsilon-2\,k b ^{3}\epsilon+2\,a a_{t} b_{t}  b ^{2}\epsilon+a\sigma_{l} ba_{l}-2\,a b_{l} \sigma a_{l}+2\,b\sigma aa_{ll}-2\,\epsilon\, b ^{2} b_{tt}  a ^{2}-2\,b\sigma a_{l} ^{2})^{2}}{2\sigma^{2}b^{6}a^{4}} \label{cc}
\end{equation}
where a subscript denotes differentiation. The calculation proceeds in about $\frac{1}{5}$ second on a contemporary PC under GRTensorII $^2$. It turns out that for metrics of the form (\ref{metric}), prior to the vacuum requirement,  $R_{AB}R^{AB}$ is cumbersome. Even the Ricci scalar turns out to be larger than the Weyl invariant calculated here. Under the restrictions (11) and (12) in Fukui \textit{et al }$^1$ (\ref{cc}) reduces to their equation (13).

\section*{{\large ACKNOWLEDGMENTS}}
This work was supported by a grant from the Natural Sciences and Engineering Research Council of Canada.  

\section*{}
$^{~1}$~T.~Fukui, S.~Seahra and P.~S.~Wesson, gr-qc/0105112.\\
$^{~2}$Portions of this work were made possible by use of \textit{GRTensorII}. This is a package which runs within Maple. It is entirely distinct from packages distributed with Maple and must be obtained independently. The GRTensorII software and documentation is distributed freely on the
World-Wide-Web from the address {\tt http://grtensor.phy.queensu.ca}.
Further details concerning the calculation reported here can be found at {\tt http://grtensor.phy.queensu.ca/five5}.

\end{document}